\documentclass[journal=jctcce,manuscript=article,]{achemso}

\usepackage[version=3]{mhchem} 

\usepackage{longtable,booktabs}

\usepackage{graphicx,grffile}
\makeatletter
\def\maxwidth{\ifdim\Gin@nat@width>\linewidth\linewidth\else\Gin@nat@width\fi}
\def\maxheight{\ifdim\Gin@nat@height>\textheight\textheight\else\Gin@nat@height\fi}
\makeatother
\setkeys{Gin}{width=\maxwidth,height=\maxheight,keepaspectratio}

\IfFileExists{parskip.sty}{%
\usepackage{parskip}
}{
\setlength{\parindent}{0pt}
\setlength{\parskip}{6pt plus 2pt minus 1pt}
}

\providecommand{\tightlist}{%
  \setlength{\itemsep}{0pt}\setlength{\parskip}{0pt}}

\usepackage{caption} \usepackage{subcaption}
\usepackage[unicode={true}]{hyperref}

\author{Sky(Yixiang) Zhang}

\affiliation[DOC, Tsinghua University]{Department of Chemistry, Tsinghua University, Beijing, 100084, P.R.
China}

\author{Hai Xiao}

\email{haixiao@tsinghua.edu.cn}

\affiliation[DOC, Tsinghua University]{Department of Chemistry, Tsinghua University, Beijing, 100084, P.R.
China}

\author{Jun Li}

\email{junli@tsinghua.edu.cn}

\affiliation[DOC, Tsinghua University]{Department of Chemistry, Tsinghua University, Beijing, 100084, P.R.
China}

\title{An Efficient Strategy to Generate Atom Connecting Positions}

\begin{document}
%
\begin{abstract}
Atom connecting positions(ACPs) are positions where an atom is
connecting to another one or a few atoms, which is needed when
constructing final state used in chain-of-state(CoS) methods for
transition state(TS) locating and minimum energy path(MEP) searching,
especially with bond formation. However, ACPs are generated with
chemical insight and experience, which is not only low efficient and
time wasting, but the manually generated structure may be far from the
optimized one. A efficient method is presented here for generating ACPs
which is based on spherical optimization and VSEPR theory without manual
interfering. Several examples are testified to prove the efficiency and
robustness of the method.
\end{abstract}

\section{INTRODUCTION}

Initial and final state should be prepared before doing transition
state(TS) locating and minimum energy path(MEP) searching using
chain-of-state(CoS) methods like nudge elastic band(NEB) method
\cite{Henkelman2000Improved, NEB-book, NEB2012, Sheppard2008Optimization,
Sheppard2011Paths, Sheppard2012A}. Generally speaking, with a bond
formation described in the MEP, the initial state is obtainable, whereas
the final state is usually unknown and should be constructed
artificially. Since there is bond reforming and atom transferring,
determining suitable positions for the atom to be positioned is really
important. These positions are defined as atom connecting
positions(ACPs). However, these positions are generally determined
manually based on chemical insight and experience, because these
positions correlate with geometry structure as well as electronic
structure.

Valence shell electron pair repulsion(VSEPR) theory is a good start,
since the ACPs determined by chemical insight are generally based on the
repulsion of connected atoms. When one atom is connecting to the
destination atom, suiting the destination atom to one of the VSEPR
shapes like Linear, Tetrahedral, Pentagonal planar or Octahedral is a
possible option. ACPs could then be generated with certain VSEPR shape.

But direct usage of VSEPR shape model is not only limited and
complicated, but also would fail under some situations. Firstly, there
are too many VSEPR shape models and
variations\cite{mackay2017introduction}. especially when distortion
exists. Next, VSEPR shape model can only be used when the transferred
atom connect to another atom, but it could also connect with multiple
ones simultaneously. Besides, VSEPR would fail when the atoms are not
connected by chemical bond, like metal surface and cluster, in which
atoms are packing together. Choosing connected atoms for suiting the
shape is very important for VSEPR, but it's difficult to classify two
atoms to be connected or not when they are weakly connected, and
different choice may give out different result. All these drawbacks may
downgrade the robustness of the method. Therefore, we need to purpose a
general, efficient and robust method for determining ACPs connecting
with not only specific atom as a center, but a virtual center like the
mid-point of two atoms or the center of a benzene ring. The center is
defined as kernel.

By rethinking the motivation of VSEPR theory, we found that the main
idea is to find out the low-repulsion positions around the spherical
surface centered at kernel. Therefore, we believe the ACPs are actually
the collection of low-repulsion positions. This leads us to a
constrained optimization problem. By optimizing all the positions on the
spherical surface centered at the kernel, we may get all ACPs. By
introducing spherical optimization(SOPT) method
\cite{Abashkin1994Transition}, we formulated the framework of
VSEPR-SOPT, which turns the chemical insight into a robust and efficient
algorithm.

\section{METHODS}

\subsection{Spherical Optimization}

The detail of spherical optimization(SOPT) has been described by
Abashkin, Y. and Russo, N. \cite{Abashkin1994Transition}, so it will be
introduced briefly. Considering two atom system with \(N\) atoms and
noted as \(x\) and \(y\) with root-mean-square deviation of distance as
\(R\), the constrained optimization problem is described as

\begin{align}
\begin{split}
&\min_{\vec{x}} E   =    E(x_1, x_2, ..., x_{n})\\
&\mathrm{s.t.}\qquad (x_1 - y_1)^2 + (x_2 - y_2)^2 + ... + (x_{n}-y_{n})^2 = R^2
\end{split}
\end{align}

where \(n=3N\) is the total degrees of freedom. \(x\) and \(y\) are also
named as target and anchor, since \(x\) will be optimized and \(y\) is
fixed during the optimization. The energy function is rewritten by
including the constraint and eliminate \(x_n\),

\begin{align}
E' = E(x_1, x_2, ..., x_{n-1}, f(x_1, x_2, ..., x_{n-1}, R))
\end{align}

and the force is rewritten as

\begin{align}
\begin{split}
F'_i = -\frac{\partial E'}{\partial x_i} =& -(\frac{\partial E}{\partial x_i} +
\frac{\partial E}{\partial x_{n}} \frac{\partial x_n}{\partial x_i})\\
     =& F_i - F_n \frac{x_i-y_i}{x_n-y_n}
\end{split}
\label{eq:spherical-force}
\end{align}

where \(F\) is the force obtained from electronic structure calculation,
and \(F'\) is a \(n-1\) with \(x_n\) eliminated. Thus, we convert this
particular constrained optimization with \(n\) variables to a regular
optimization problem with \(n-1\) variables. More details of the
algorithm are neglected and can be found elsewhere.

\subsection{VSEPR-SOPT Model}

Combining SOPT with VSEPR theory is tricky, since the target and anchor
system are not real system, and the energy function remains unknown.

Firstly, let's consider locating one ACP on the sphere centered at
kernel. Locating multiple ones can be regarded as locating ACP one after
another. Positions on the sphere centered at kernel with radius \(R=L\)
is the collection of possible positions. Here \(L\) is the length
between kernel and the position on the sphere, which is given by user or
get from database. So the target system is constructed with a pseudo
atom positioned on the sphere, and the anchor system contains only the
position of kernel.

The energy of the system is to describe repulsion between the pseudo
atom and any other atoms in the target system. Minimizing the energy
will give out the ACP directly. But the formula of energy is hard to
determine since the system is not real, and VSEPR repulsion has not been
formulated. Besides, we expect the calculation to be as fast as possible
without introducing too much parameters. Inspired by Lennard-Jones
potential, we believe the repulsion energy between a pair of atoms can
be formulated as \(\frac{1}{|\vec{r}|^n}\), where \(n\) is a constant
number. It turned out \(n=4\) will give out the best results. For
molecular system, the formula of energy is

\begin{align}
\begin{split}
E(\vec{x}) = \sum_{l=1}^N \frac{1}{|\vec{r_l} - \vec{x}|^{4}}
\end{split}
\end{align}

where \(N\) is the total number of atoms in real system, \(\vec{x}\) is
the position of pseudo atom and \(\vec{r_l}\) is the position of
\(l^{th}\) atom in real system. It's worth mentioning that no external
parameter is introduced in the function. It seems weird, but the results
turned out to be good enough.

For periodic system, the contribution of replicas should be counted as
well. Therefore, another 8 neighbor cells are introduced. The energy
function is

\begin{align}
\begin{split}
E(\vec{x}) = \sum_{i,j,k=-1}^{1}\sum_{l=1}^N \frac{1}{|\vec{r_l} +i\vec{a}+j\vec{b}+k\vec{c} - \vec{x}|^{4}}
\end{split}
\end{align}

where \(\vec{a}, \vec{b}, \vec{c}\) are unit vectors of the cell.

For \(M\) ACPs, we just need to construct the target system contains
\(M\) pseudo atoms, and duplicate kernel \(M\) times in anchor system,
adjust the radius of the sphere \(R=L\sqrt{M}\). Interactions between
pseudo atoms should be added to the energy function

\begin{align}
\begin{split}
E(\vec{x_1}, \vec{x_2}, ..., \vec{x_M}) = \sum_{i=1}^N\sum_{j=1}^M \frac{1}{|\vec{r_i}
- \vec{x_j}|^{4}} + \sum_{i, j=1, i\neq j}^M \frac{1}{|\vec{x_i} - \vec{x_j}|^{4}}
\end{split}
\label{eq:potential}
\end{align}

The rest part remains the same. And for periodic system with multiple
ACPs, the formula will be modified as presented above. With the formula
of energy function, SOPT now is able to be utilized for finding ACPs.
Since the simplicity of the procedure, the execution of the algorithm is
really fast.

\subsection{Position Sampling on Spherical Surface}

In order to acquire all ACPs on the spherical surface, we need to sample
all positions on the surface. Therefore, a sampling algorithm should be
introduced. Using polar coordinate system and sampling \(\theta\) and
\(\phi\) uniformly is simple and convenient, but the sampling density
around the pole is much higher than that around the equator. However, a
uniform sampling on the spherical surface method is expected. Here we
use a method introduced by Cory Simon\cite{sampling-sphere}. By
generating three standard normally distributed numbers \(X\), \(Y\),
\(Z\) to form a vector \(V=[X, Y, Z]\) and normalize it, the vector is
uniformly distributed on the surface of sphere. Normally, sampling 100
times is adequate for covering the entire surface, but the number can be
increased if necessary.

\subsection{Flowchart of VSEPR-SOPT}

We provide the flowchart for the algorithm of VSEPR-SOPT below,

\begin{enumerate}
\def\labelenumi{\arabic{enumi}.}
\tightlist
\item
  Input the system, kernel position, number of ACPs needed(\(N_n\)).
\item
  Add a pseudo atom to the system and construct energy function.
\item
  Sampling on the sphere \(N_s\) times with the algorithm described
  above, optimize the pseudo atoms system with the potential described
  in Eq.\ref{eq:potential} with SOPT.
\item
  Repeat step 2-3 \(N_n\) times to get all ACPs needed.
\end{enumerate}

\section{RESULTS}

Here we use the notation A@B to represent system A with B as the kernel,
e.g. \ce{CH2O}@C means a molecular system \ce{CH2O}, with the carbon
atom as the kernel. It's worth mentioning that the center of a ring can
be kernel as well. The kernel will be noted as circle in these
situations. In the following, we tested several kinds of system,
including small molecules (\ce{CH4}, \ce{CH2O}), clusters (\ce{Au20},
\ce{C60}), complicated molecular system(complicated \ce{Au6} cluster)
and heterogeneous system(\ce{CeO2}-based single atom catalyst(SAC)). In
the related figures, ACPs are labeled with pseudo atoms in violet, and
only kernel(or kernel related) atoms and pseudo atoms are in ball-stick
format and other atoms are in line format. The B3LYP functional with
default parameters in Gaussian 09\cite{g09} are used in all DFT
calculations and geometry optimization if not specified. 6-31G(d) basis
sets are used for C, O, H and LANL2DZ are used for Au, Fe. Difference of
angle between VSEPR-SOPT and DFT geometry optimization are calculated
with other atoms aligned, and the numbers are shown in
Table.\ref{table:angle-diff}. This proves the validity of our method.

\begin{longtable}[]{@{}lr@{}}
\caption{Angle Difference between VSEPR-SOPT and DFT Optimization
\label{table:angle-diff}}\tabularnewline
\toprule
Item & Angle Difference\tabularnewline
\midrule
\endfirsthead
\toprule
Item & Angle Difference\tabularnewline
\midrule
\endhead
\ce{Au20}@Au\textsubscript{a} & \(0.4^\circ\)\tabularnewline
\ce{Au20}@Au\textsubscript{e} & \(13.43^\circ\)\tabularnewline
\ce{Au20}@Au\textsubscript{c} & \(2.24^\circ\)\tabularnewline
CNT@circle & \(2.48^\circ\)\tabularnewline
\ce{C60}@circle & \(8.13^\circ\)\tabularnewline
\ce{Au1}/\ce{CeO2} & \(13.19^\circ\)\tabularnewline
\bottomrule
\end{longtable}

\subsection{\texorpdfstring{\ce{CH4} and \ce{CH2O}
System}{ and  System}}

\begin{figure}[htbp]

\centering
\begin{subfigure}[t]{0.49\linewidth}
\centering
\includegraphics[width=0.5\linewidth]{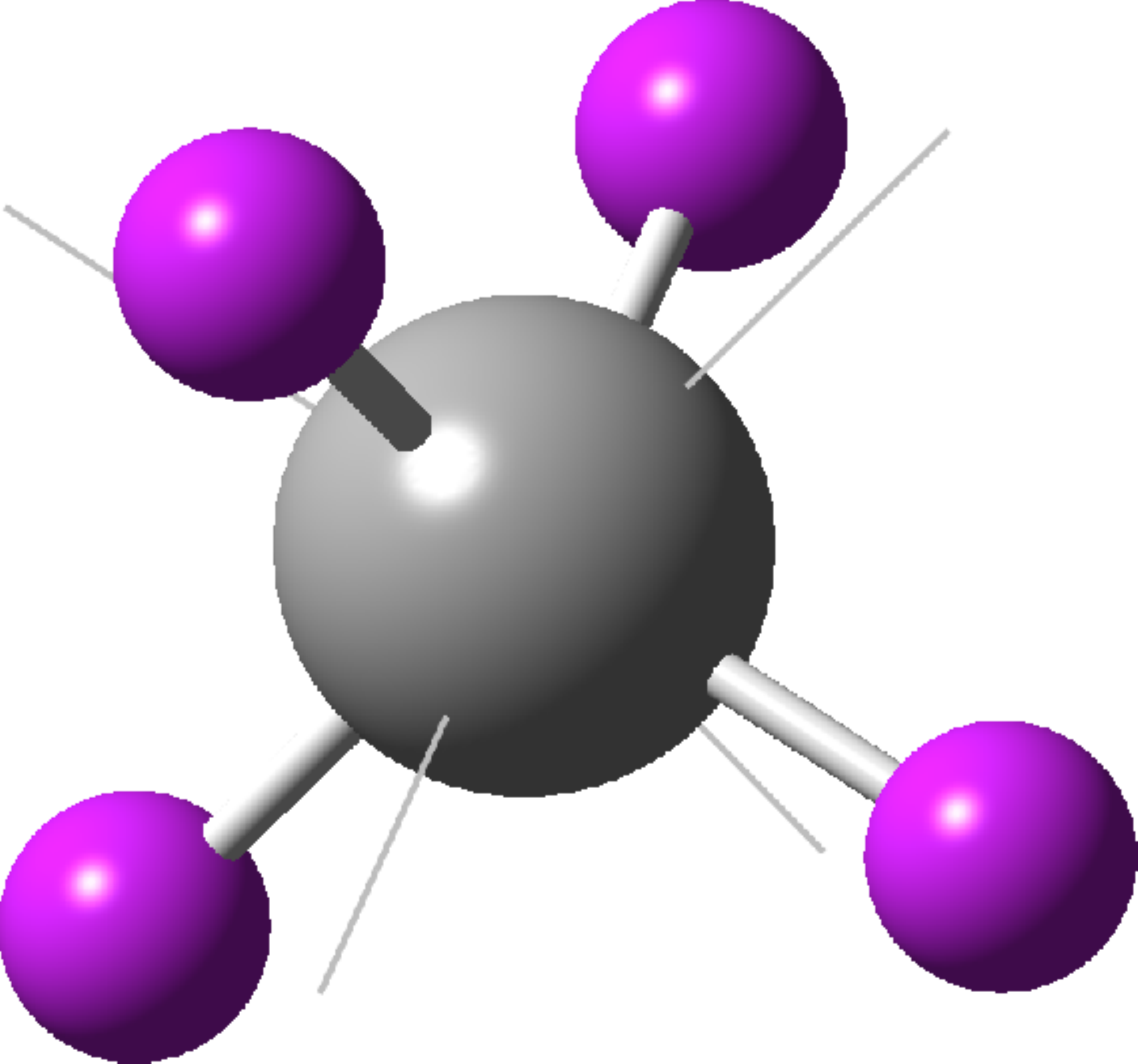}
\caption{\ce{CH4}@C}
\label{fig:ch4-c}
\end{subfigure}
\begin{subfigure}[t]{0.49\linewidth}
\centering
\includegraphics[width=0.5\linewidth]{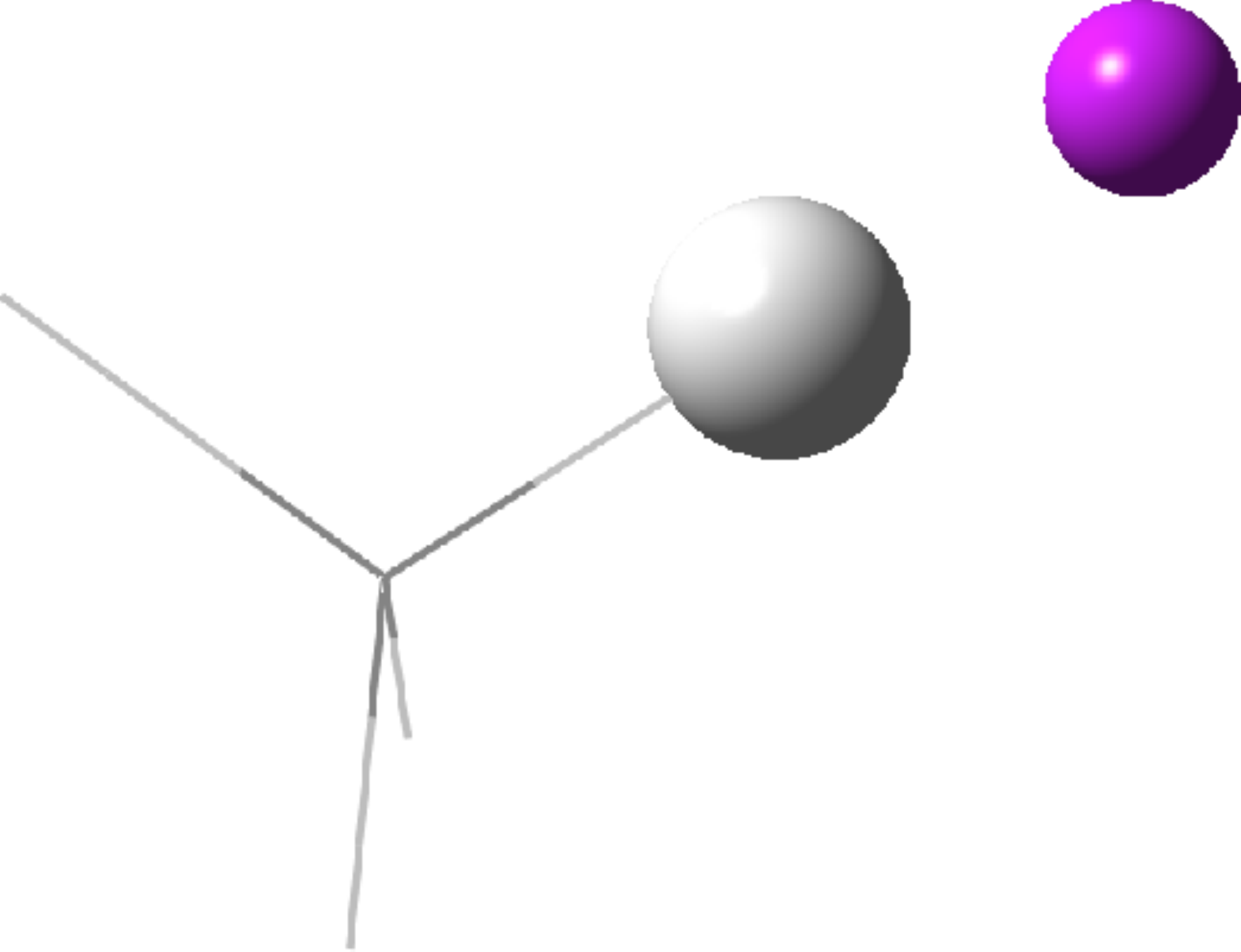}
\caption{\ce{CH4}@C}
\label{fig:ch4-h}
\end{subfigure}

\caption{\ce{CH4} ACPs with C, H as kernel}
\end{figure}

\begin{figure}[htbp]

\centering
\begin{subfigure}[t]{0.49\linewidth}
\centering
\includegraphics[width=0.5\linewidth]{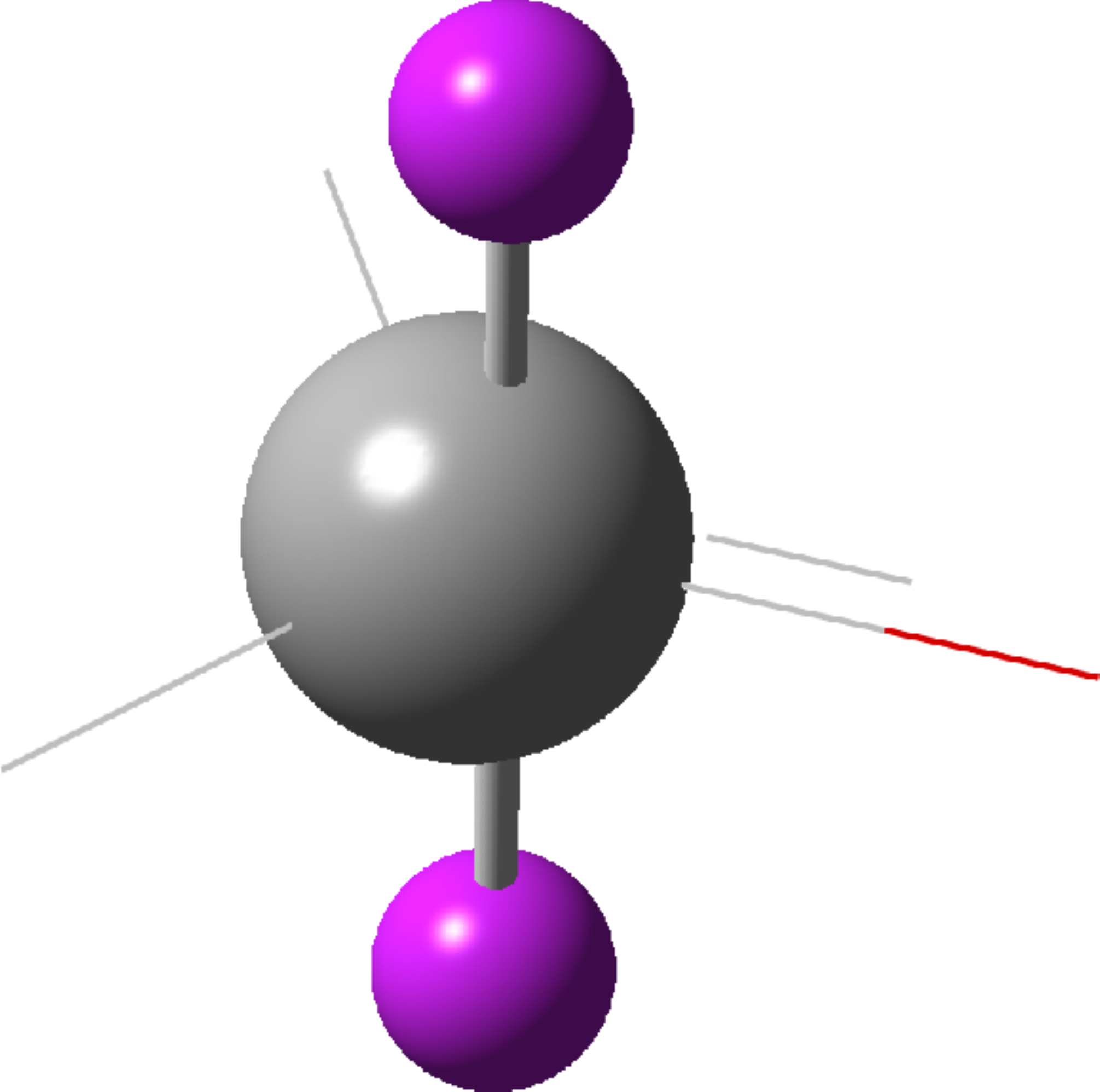}
\caption{\ce{CH2O}@C}
\label{fig:ch2o-c}
\end{subfigure}
\begin{subfigure}[t]{0.49\linewidth}
\centering
\includegraphics[width=0.5\linewidth]{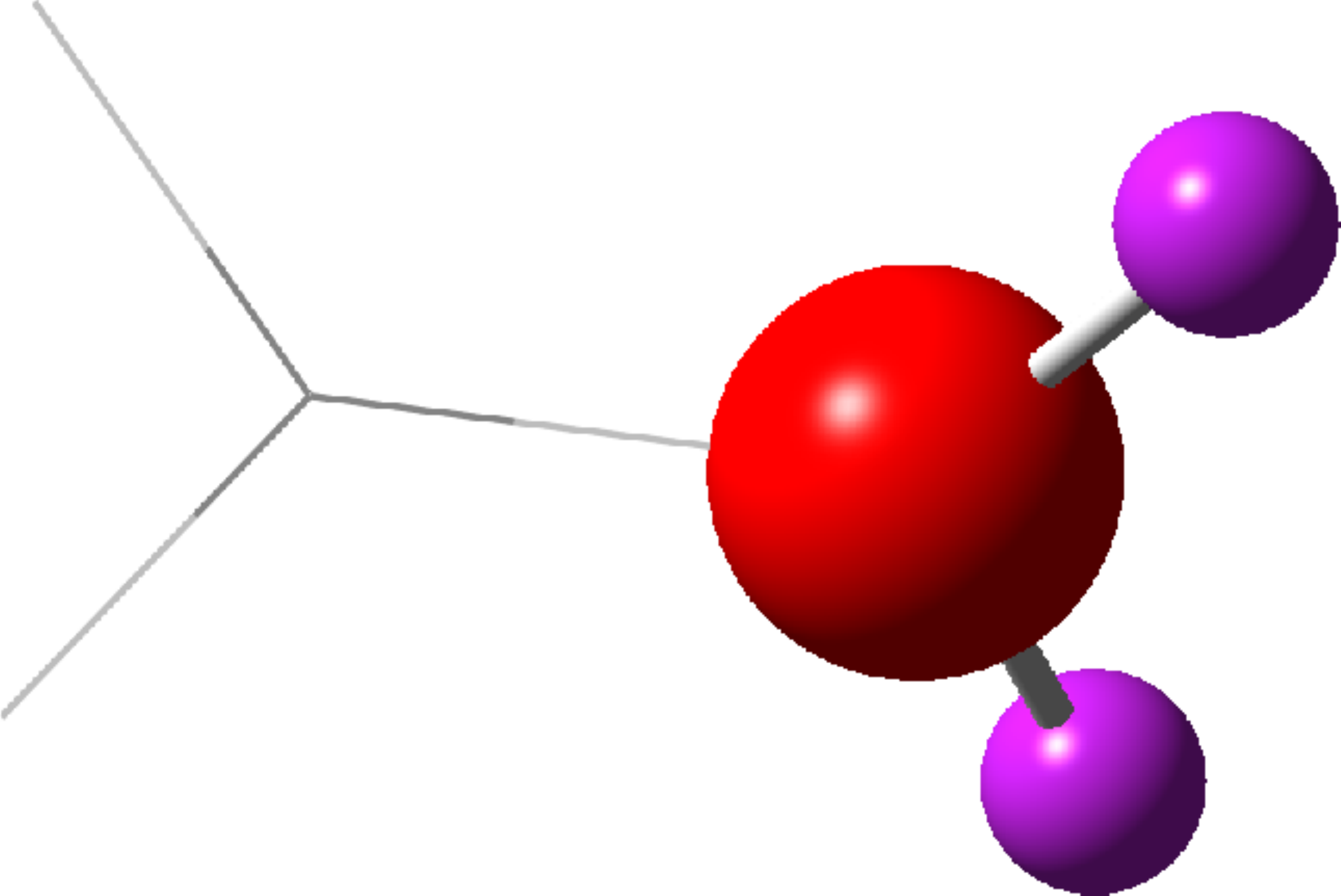}
\caption{\ce{CH2O}@O}
\label{fig:ch2o-o}
\end{subfigure}

\caption{\ce{CH2O} ACPs with C, O as kernel}
\end{figure}

VSEPR-SOPT results of \ce{CH4}@C, \ce{CH4}@H are shown in
Fig.\ref{fig:ch4-c} and \ref{fig:ch4-h}. For \ce{CH4}@C, all 4 face
centers of the tetrahedral are acquired, and for \ce{CH4}@H, linear
shape is implied and the other end is acquired as ACP. Results of
\ce{CH2O}@C, \ce{CH2O}@O are shown in Fig.\ref{fig:ch2o-c} and
\ref{fig:ch2o-o}. For \ce{CH2O}@C, a trigonal bipyramidal is displayed
and the positions off the \ce{CH2} plane are acquired. However, for
\ce{CH2O}@O, the lone-pairs' positions acquired violates the chemical
rule, since we expect the lone-pairs to be in the plane of \ce{CH2O},
but the positions given by VSEPR-SOPT are off the plane. This is due to
the free sampling of the spherical surface and the absence of electronic
structure. But the result is still acceptable. The results shown here
satisfy chemical insights in general, and the weakness is endurable.

\subsection{\texorpdfstring{\ce{Au20} Cluster System}{ Cluster System}}

\begin{figure}[htbp]

\centering
\begin{subfigure}[t]{0.3\linewidth}
\centering
\includegraphics[width=0.5\linewidth]{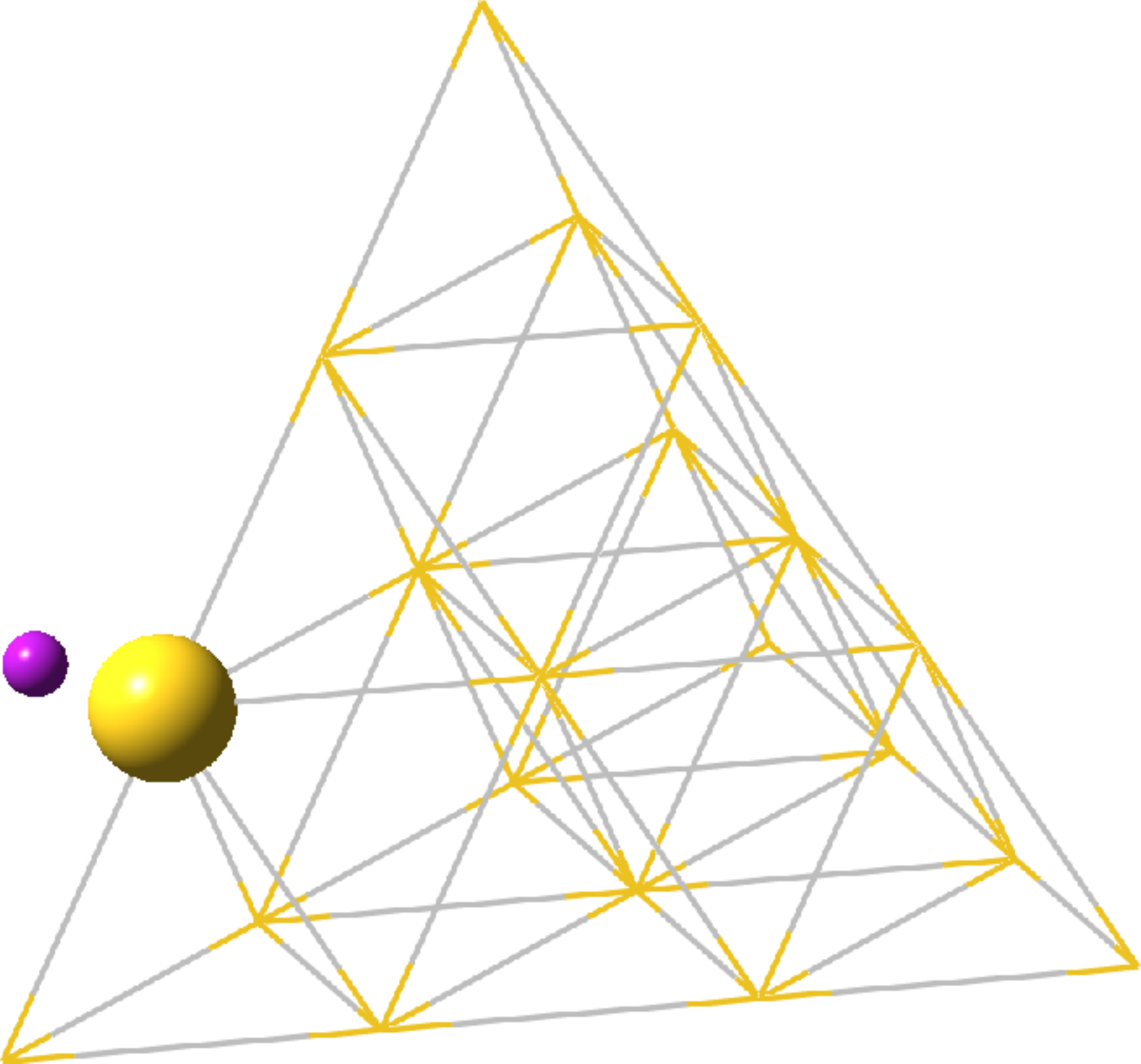}
\caption{\ce{Au20}@Au\textsubscript{e}}
\label{fig:au20-e}
\end{subfigure}
\begin{subfigure}[t]{0.3\linewidth}
\centering
\includegraphics[width=0.5\linewidth]{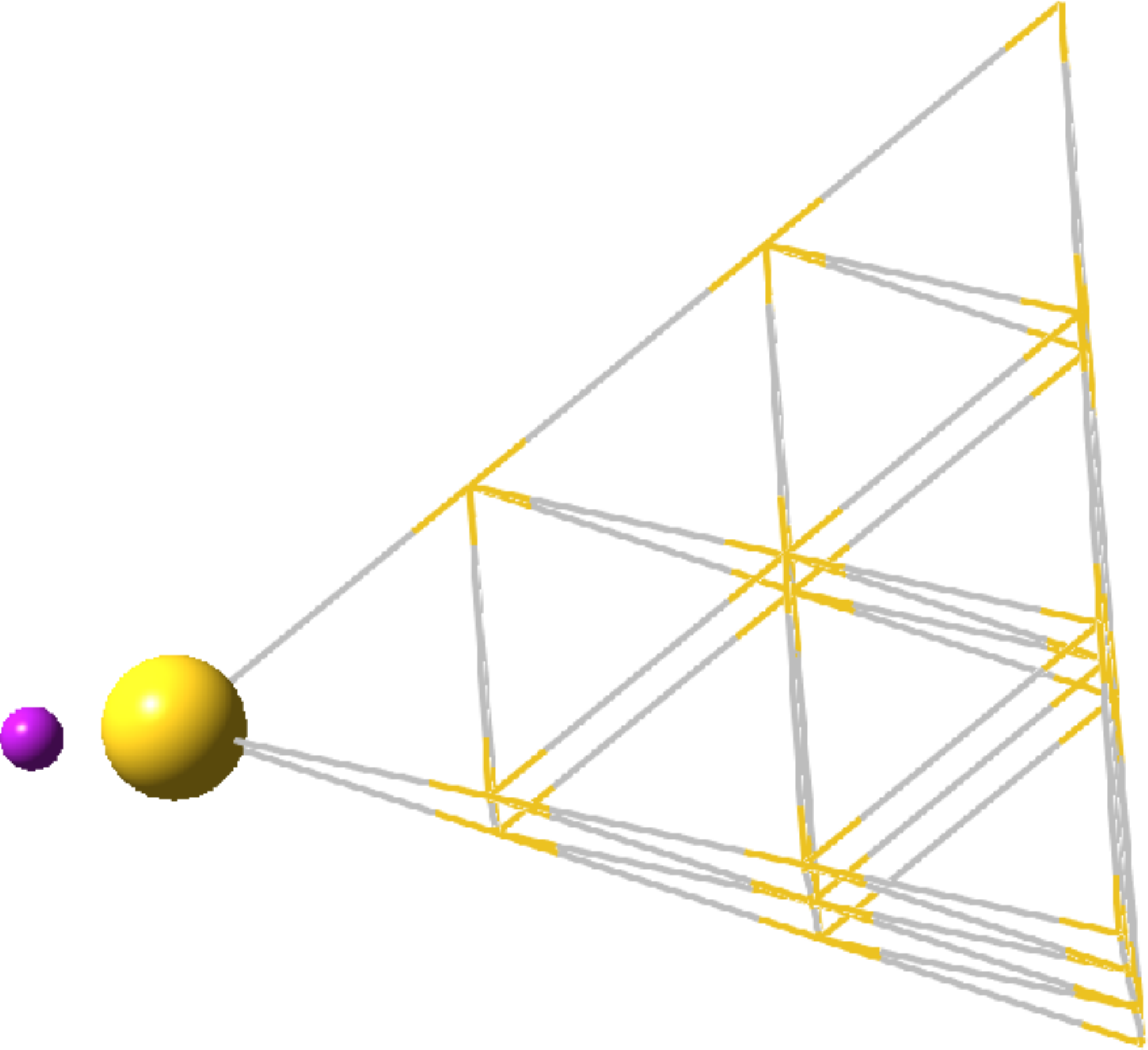}
\caption{\ce{Au20}@Au\textsubscript{a}}
\label{fig:au20-a}
\end{subfigure}
\begin{subfigure}[t]{0.3\linewidth}
\centering
\includegraphics[width=0.5\linewidth]{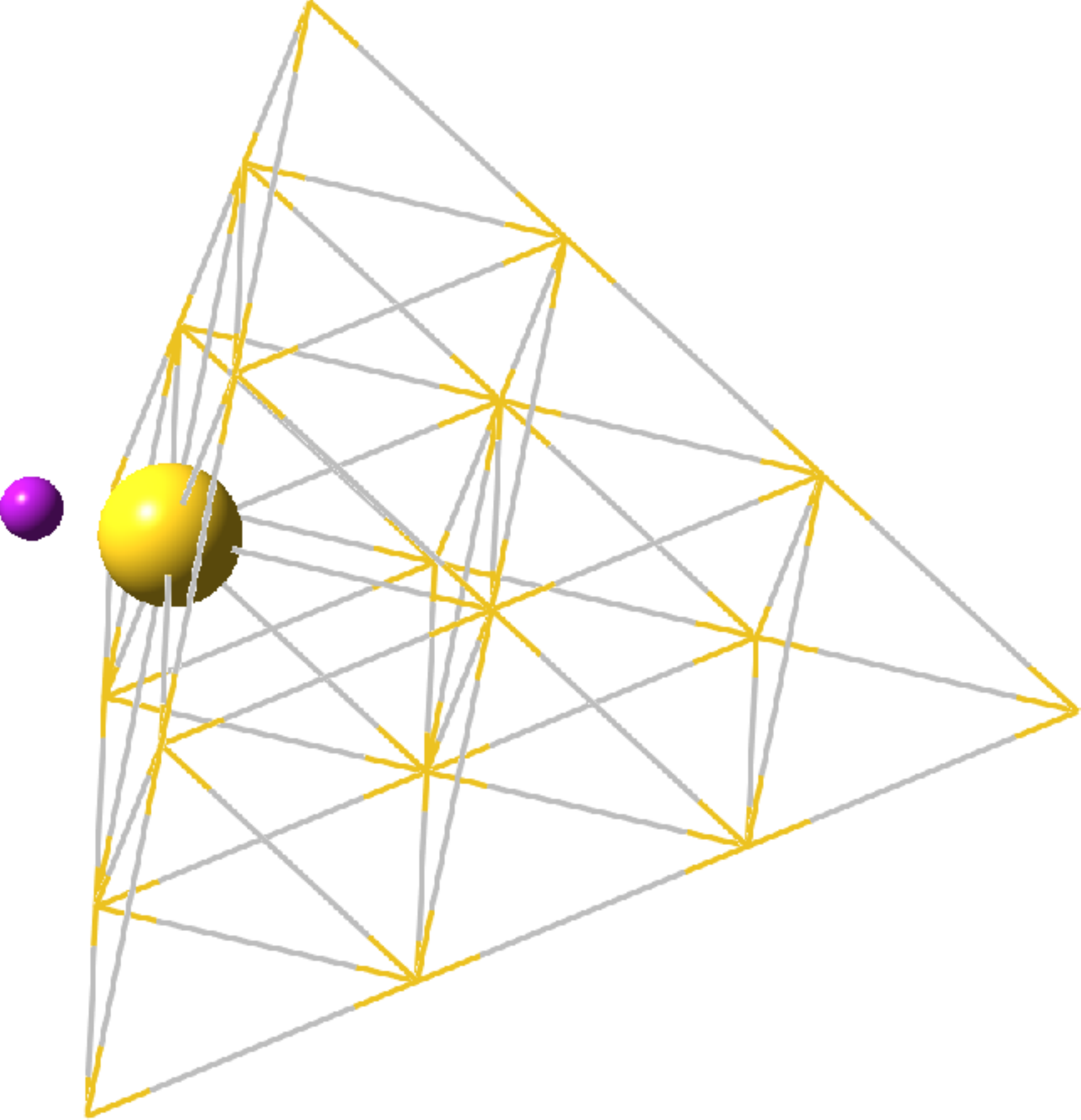}
\caption{\ce{Au20}@Au\textsubscript{c}}
\label{fig:au20-c}
\end{subfigure}

\caption{\ce{Au20} ACPs}
\end{figure}

\begin{figure}[htbp]

\centering
\begin{subfigure}[t]{0.3\linewidth}
\centering
\includegraphics[width=0.5\linewidth]{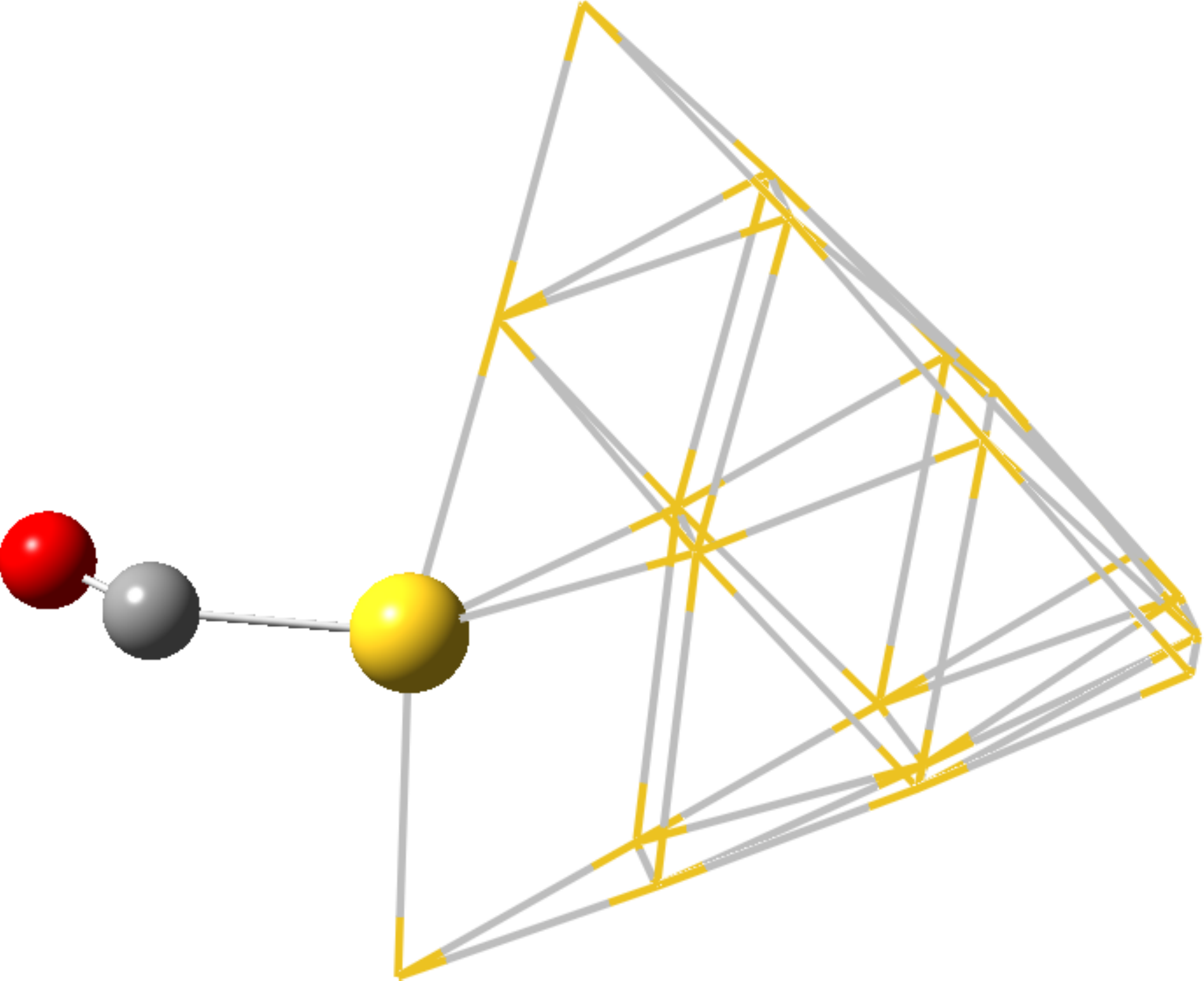}
\caption{\ce{Au20}@Au\textsubscript{e}-CO}
\label{fig:au20-e-CO}
\end{subfigure}
\begin{subfigure}[t]{0.3\linewidth}
\centering
\includegraphics[width=0.5\linewidth]{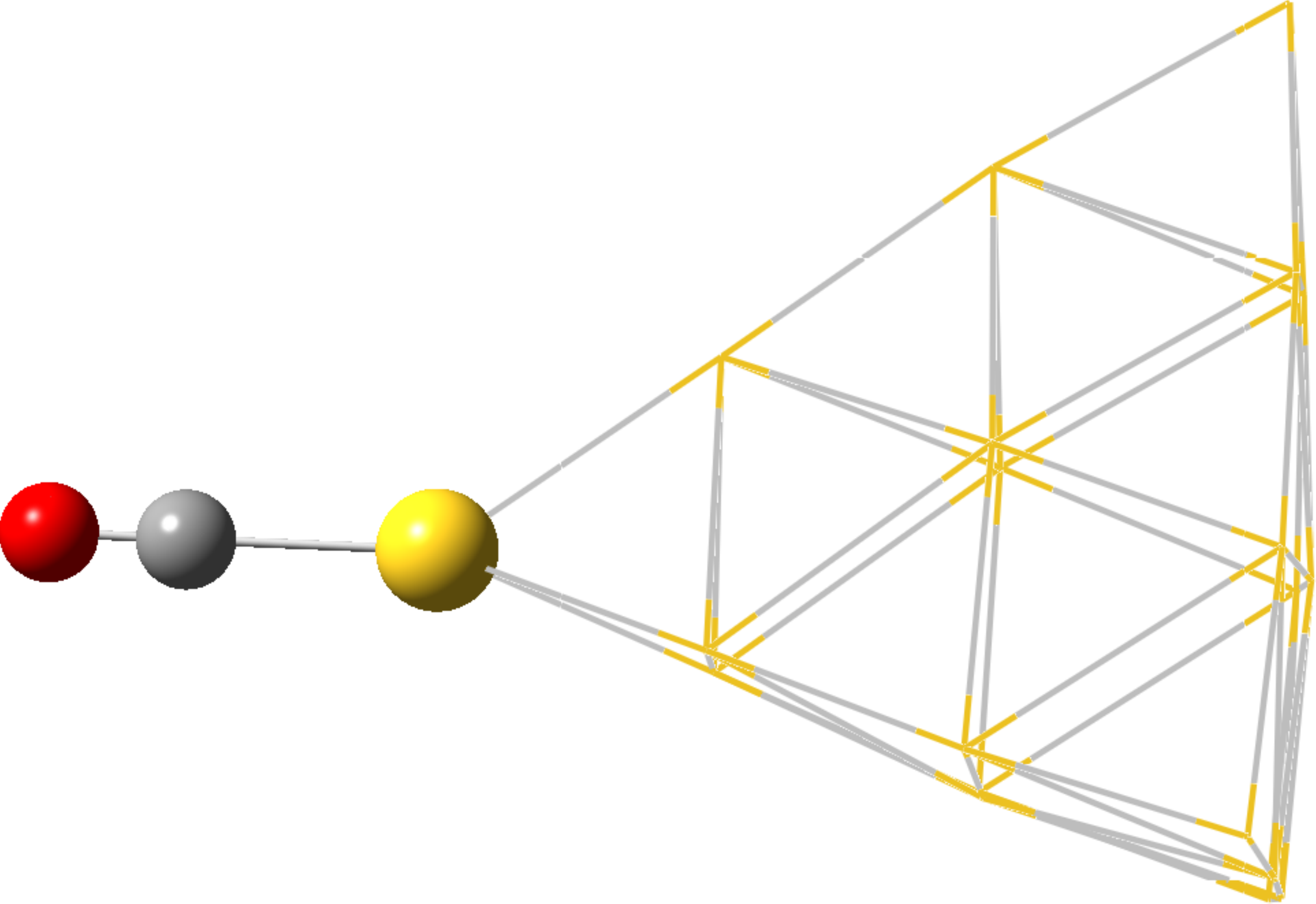}
\caption{\ce{Au20}@Au\textsubscript{a}-CO}
\label{fig:au20-a-CO}
\end{subfigure}
\begin{subfigure}[t]{0.3\linewidth}
\centering
\includegraphics[width=0.5\linewidth]{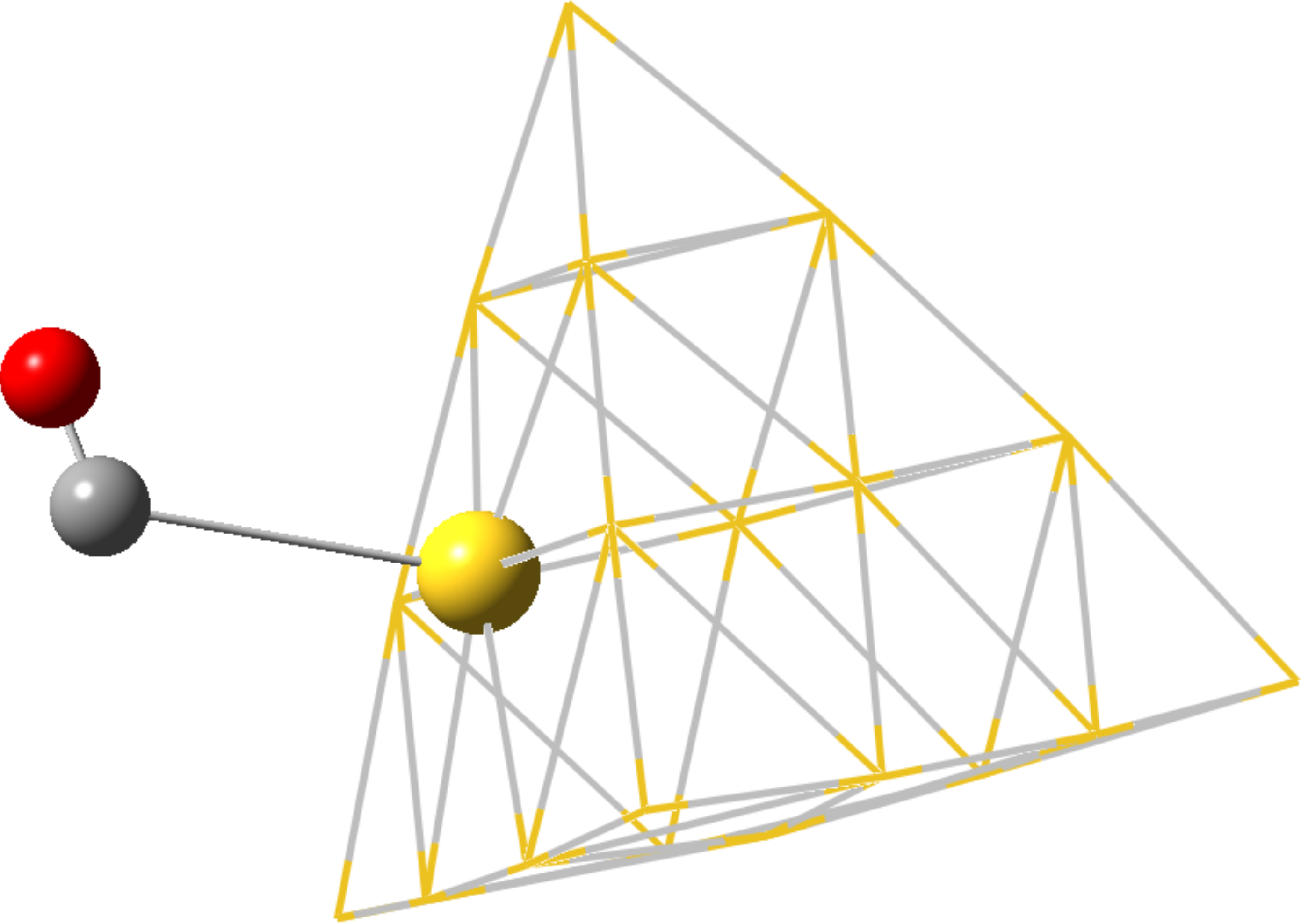}
\caption{\ce{Au20}@Au\textsubscript{c}-CO}
\label{fig:au20-c-CO}
\end{subfigure}

\caption{\ce{Au20}-CO}
\end{figure}

\ce{Au20} cluster\cite{Au20} has 3 kinds of Au: Au along the
edges(Au\textsubscript{e}), Au at the apexes(Au\textsubscript{a}) and Au
at the center of each face(Au\textsubscript{c}). ACPs for \ce{Au20} with
each kind of Au atom acquired by VSEPR-SOPT are shown in
Fig.\ref{fig:au20-e}-\ref{fig:au20-c}. For each kind of Au atom, only
one position is identified as ACP. To show the correctness of the
result, we further testified geometry structure of CO absorbed on
\ce{Au20} with DFT calculation, as shown in
Fig.\ref{fig:au20-e-CO}-\ref{fig:au20-c-CO}. The results given by DFT
match the VSEPR-SOPT results astonishingly. The angle differences are
\(0.4^\circ\), \(13.43^\circ\), \(2.24^\circ\) for
\ce{Au20}@Au\textsubscript{a}, \ce{Au20}@Au\textsubscript{e},
\ce{Au20}@Au\textsubscript{c}, respectively. Relative large angle
difference for \ce{Au20}@Au\textsubscript{e} is account for distortion
of \ce{Au20} cluster, but it is still acceptable.

However, suiting these kinds of Au atom with regular VSEPR shape model
is hard or even impossible, since these Au atoms are not connecting with
chemical bonds but packing together. It's really difficult to
determining the VSEPR shape for these Au atoms. For a
Au\textsubscript{a} atom, 3 Au atoms are coordinated, while the angles
are only \(60^\circ\) instead of \(120^\circ\) to suit a Triangle model
or \(90^\circ\) to suit a Octahedral model. For a Au\textsubscript{e}
atom, 6 Au atoms are coordinated in a strange shape which doesn't belong
to any VSEPR model. And Au\textsubscript{c} atom is coordinated with 9
Au atoms, which is far beyond the range of VSEPR shape model. This may
cause failure for regular VSEPR model, whereas the result given by
VSEPR-SOPT match our chemical insight and the reality remarkably.

\subsection{\texorpdfstring{\ce{C60} and Carbon Nanotube(CNT)
System}{ and Carbon Nanotube(CNT) System}}

For \ce{C60} and CNT, all carbon atoms are equivalent, but the ACPs can
be inside or outside the cage/tube. For simplicity, only outside of
cage/tube is testified. In these cases, center of a carbon circle can
also be kernel, and these circles are equivalent as well. So we use both
carbon and the center of circle to be the kernel.

\begin{figure}[htbp]

\centering
\begin{subfigure}[t]{0.3\linewidth}
\centering
\includegraphics[width=0.5\linewidth]{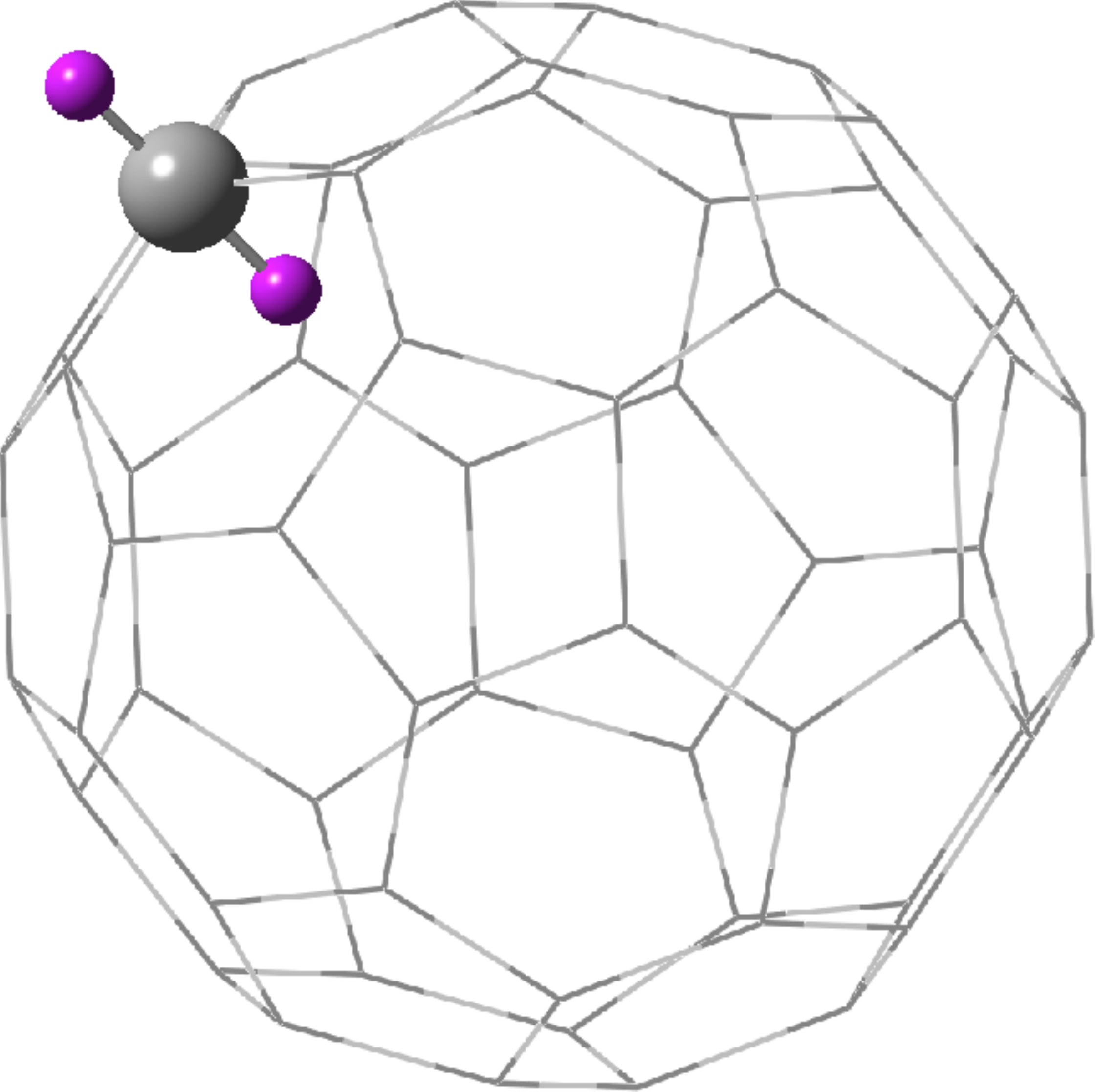}
\caption{\ce{C60}@C}
\label{fig:c60-c}
\end{subfigure}
\begin{subfigure}[t]{0.3\linewidth}
\centering
\includegraphics[width=0.5\linewidth]{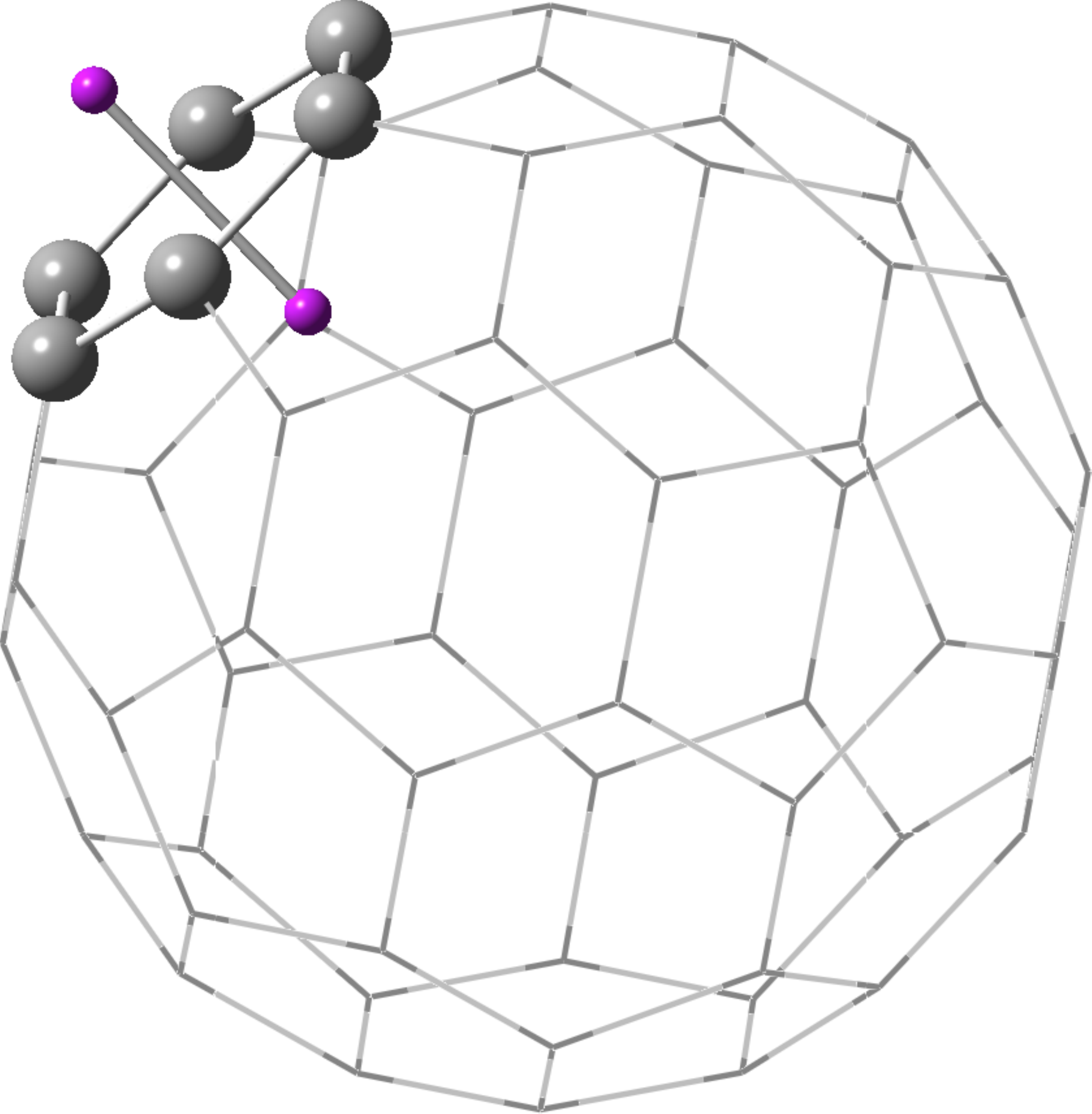}
\caption{\ce{C60}@\ce{C6}}
\label{fig:c60-c6}
\end{subfigure}
\begin{subfigure}[t]{0.3\linewidth}
\centering
\includegraphics[width=0.5\linewidth]{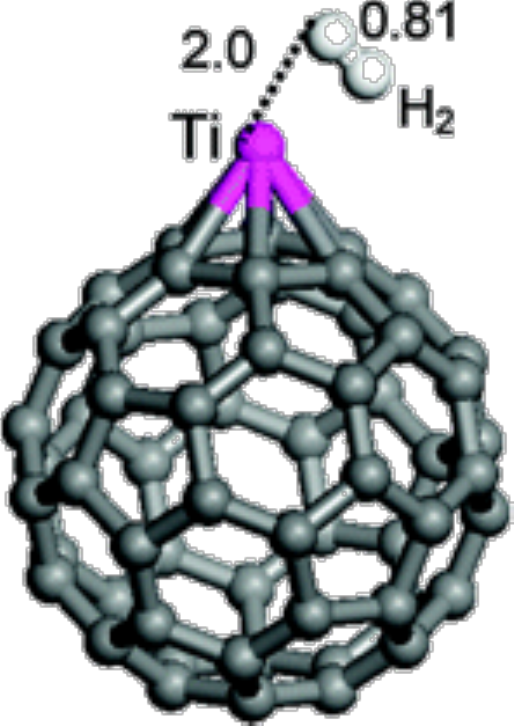}
\caption{\ce{TiC60}}
\label{fig:tic60}
\end{subfigure}

\caption{\ce{C60} system}
\end{figure}

\ce{C60} has two kinds of carbon circle, \ce{C5} circle and \ce{C6}
circle. Here we only use \ce{C6} circle and \ce{C6} could be processed
with the same strategy. Using VSEPR-SOPT, we can identify the ACPs on
\ce{C60} easily, as shown in Fig.\ref{fig:c60-c} and \ref{fig:c60-c6}.
\ce{TiC60} presented by Sun, Qiang, et al. \cite{sun2005clustering} is
optimized with PAW method implemented with VASP. The result shown in
Fig.\ref{fig:tic60} agrees with the \ce{C6} kernel situation very well
and the angle difference is about \(8^\circ\).

\begin{figure}[htbp]

\centering
\begin{subfigure}[t]{0.3\linewidth}
\centering
\includegraphics[width=0.5\linewidth]{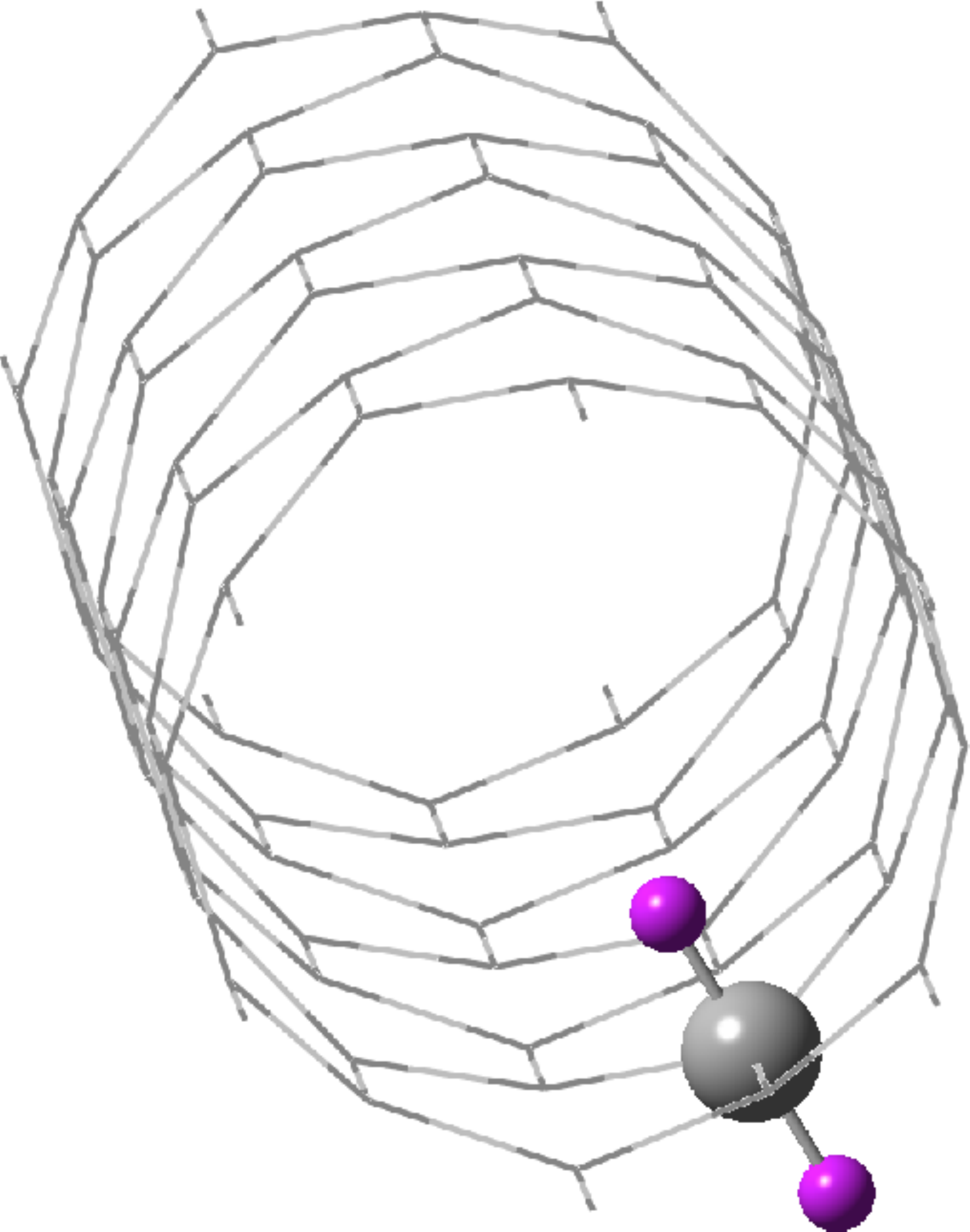}
\caption{\ce{CNT}@C}
\label{fig:cnt-c}
\end{subfigure}
\begin{subfigure}[t]{0.3\linewidth}
\centering
\includegraphics[width=0.5\linewidth]{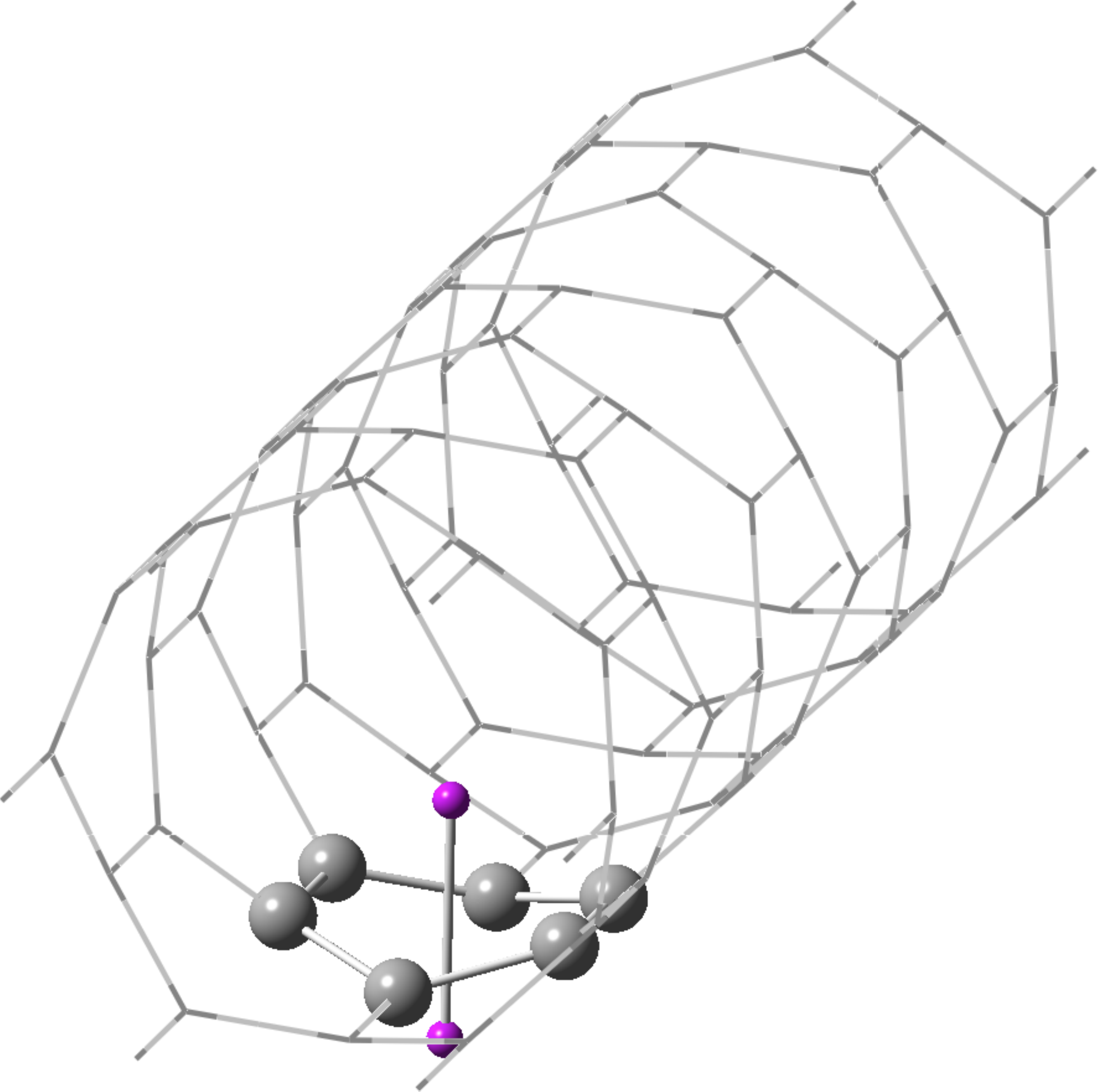}
\caption{\ce{CNT}@\ce{C6}}
\label{fig:cnt-circle}
\end{subfigure}
\begin{subfigure}[t]{0.3\linewidth}
\centering
\includegraphics[width=0.5\linewidth]{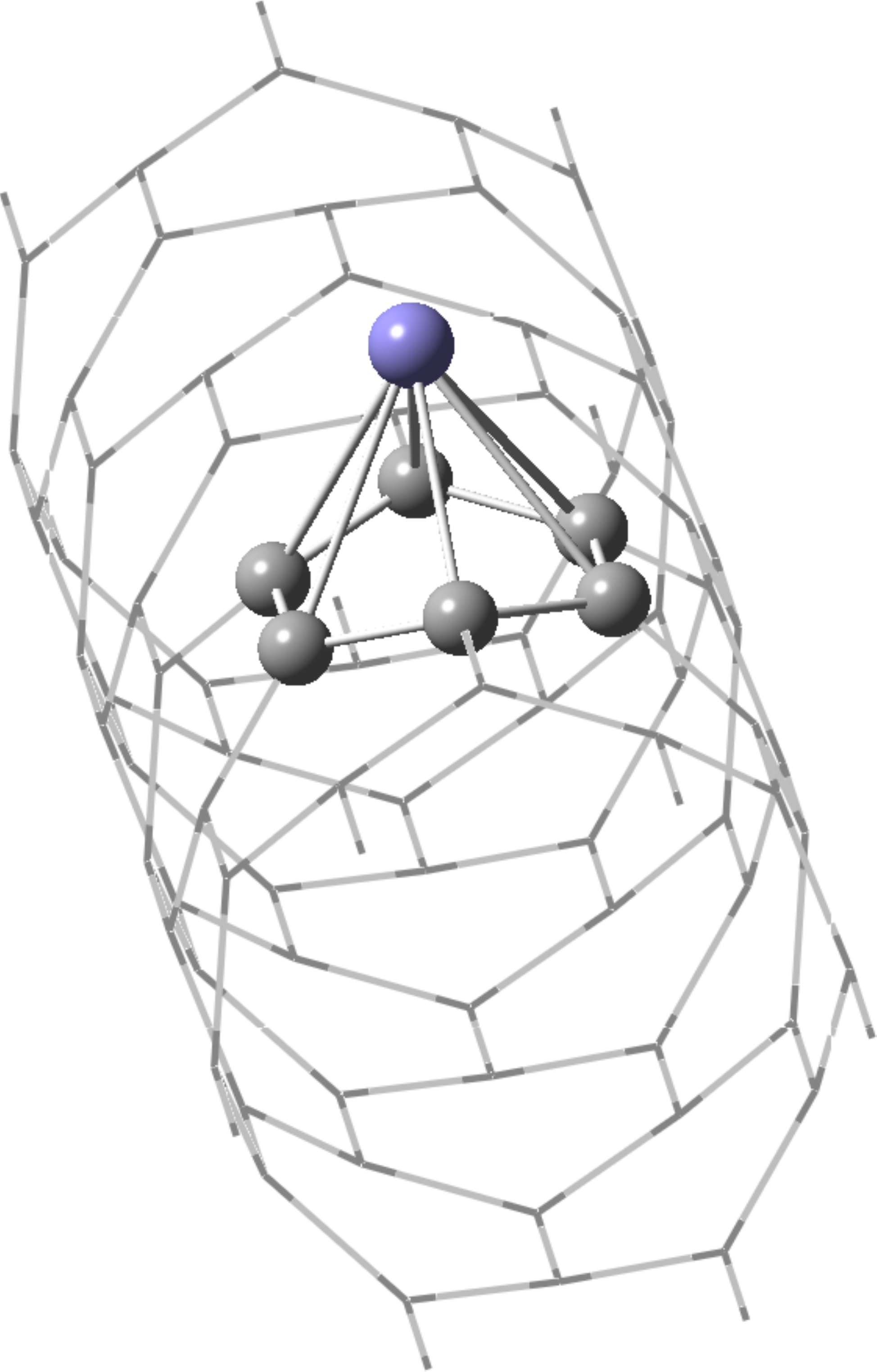}
\caption{\ce{FeCNT}}
\label{fig:cnt-Fe}
\end{subfigure}

\caption{CNT system}
\end{figure}

As for CNT, the same setups are executed. The VSEPR-SOPT results with
carbon and \ce{C6} as kernel are shown in Fig.\ref{fig:cnt-c} and
\ref{fig:cnt-circle}, Fe-CNT system is optimized with DFT and presented
in Fig.\ref{fig:cnt-Fe}. DFT result shows that Fe is just above the
center of the \ce{C6} circle, as what we acquired with VSEPR-SOPT. The
angle difference is less than \(3^\circ\), which agrees with the reality
extraordinarily.

For \ce{C60} system and CNT system, ACPs acquired by VSEPR-SOPT are not
only chemically meaningful, but also satisfy the DFT calculation
results. Whereas for regular VSEPR shape model suiting, it would be
really hard since the kernel here is not an individual atom but the
center of a carbon circle.

\subsection{\texorpdfstring{A Complicated \ce{Au6}
Cluster}{A Complicated  Cluster}}

\begin{figure}
\centering
\includegraphics[width=0.50000\textwidth]{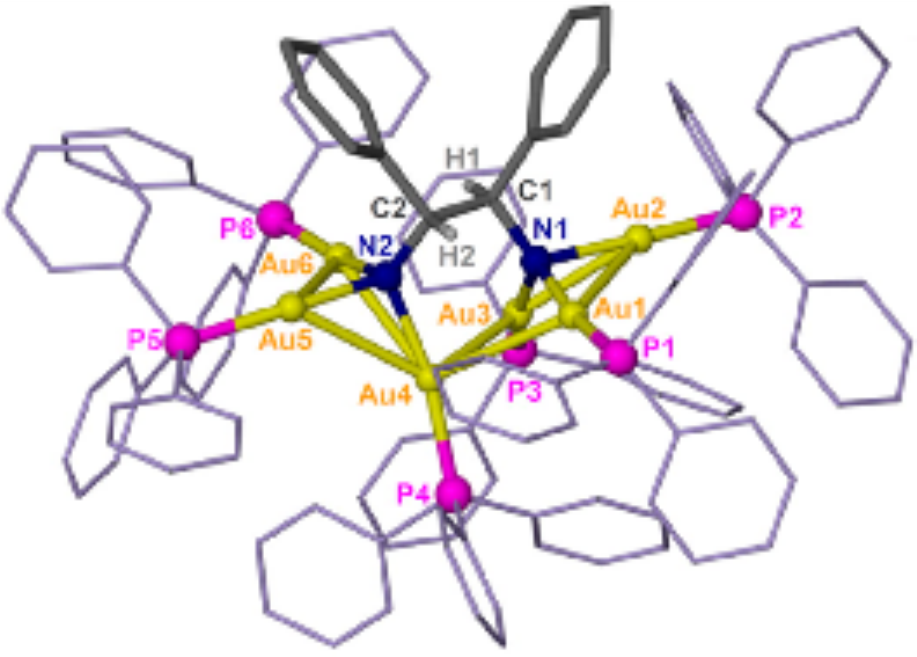}
\caption{\ce{Au6} cluster \label{fig:jacs-au6}}
\end{figure}

\begin{figure}
\centering
\includegraphics[width=0.50000\textwidth]{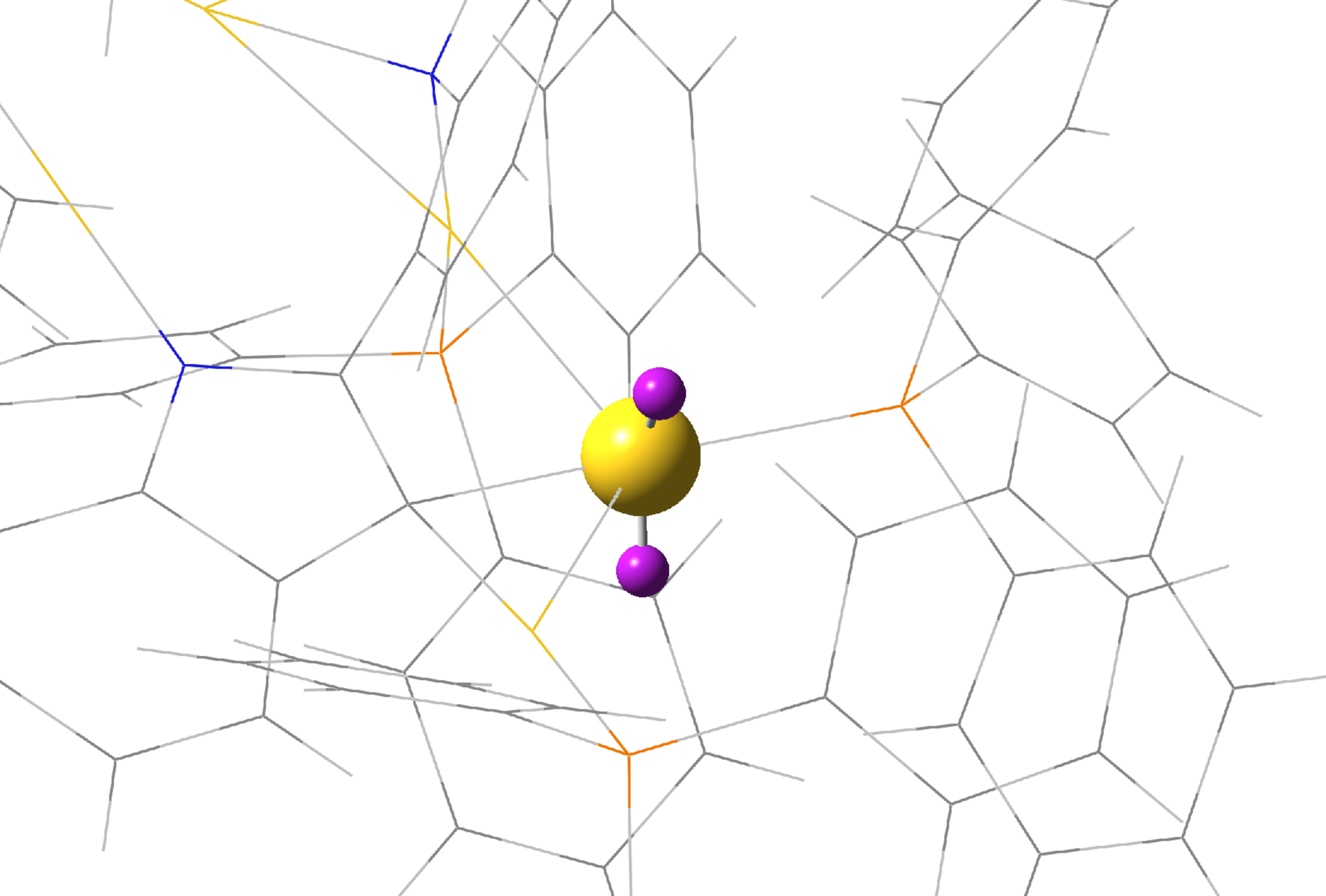}
\caption{\ce{Au6} cluster@Au by VSEPR-SOPT \label{fig:au6-cluster}}
\end{figure}

\ce{Au6} cluster\cite{Au6} is a very large cluster system, which
contains 6 equivalent Au atoms, 6 P atoms, 2 N atoms and 20 phenyl
groups(Fig.\ref{fig:jacs-au6}). Environment of the Au atom is very
complicated. One Au atom is not only connecting with one N and P, but
connecting with other Au atoms weakly, the phenyl groups around the atom
have influence as well. Therefore, it's impossible to determine the
VSEPR shape of Au atom. It's still very hard to get the positions
manually with chemical insight. However, with the help of VSEPR-SOPT,
ACPs could be acquired easily, ACPs acquired by the algorithm are shown
in Fig.\ref{fig:au6-cluster}. Noted that the VSEPR-SOPT does not need
the connection relationship at all, which is critical for regular VSEPR
shape model. This example proves the robustness of the method.

\subsection{\texorpdfstring{\ce{CeO2}-based Single Atom
Catalyst}{-based Single Atom Catalyst}}

\begin{figure}
\centering
\includegraphics{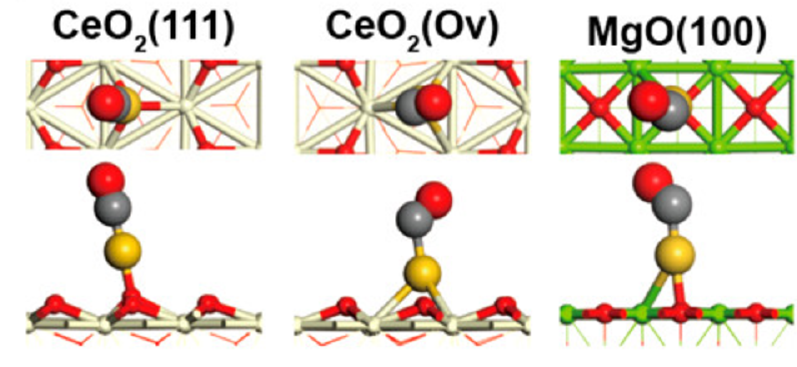}
\caption{\ce{Au1}/\ce{CeO2}-CO \label{fig:Au1CeO2-CO}}
\end{figure}

\begin{figure}
\centering
\includegraphics[width=0.50000\textwidth]{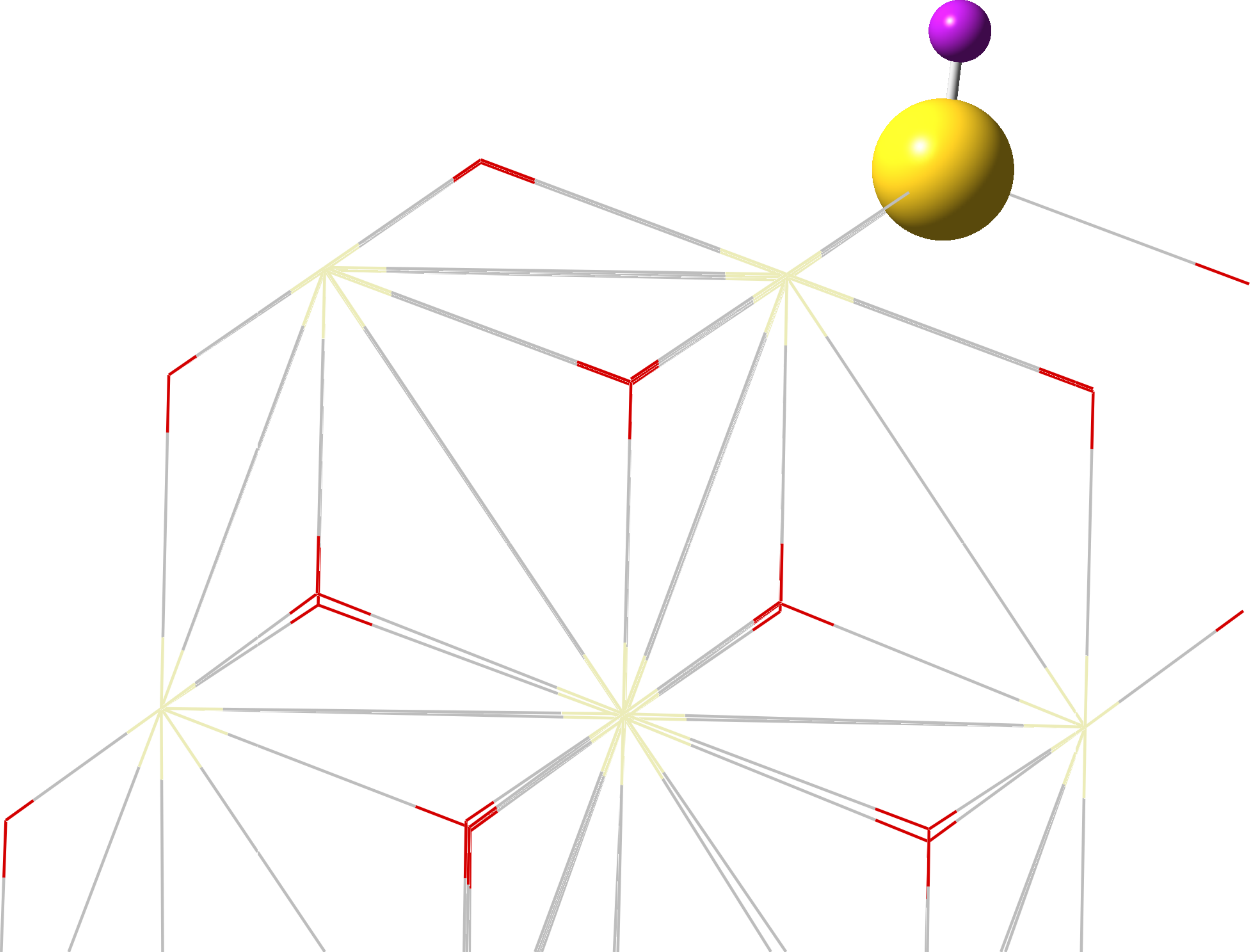}
\caption{\ce{Au1}/\ce{CeO2} ACPs \label{fig:CeO2-Au-vsepr-sopt}}
\end{figure}

\ce{CeO2}-based single atom catalyst could be used as heterogeneous
catalysis, like \ce{CO2} reduction and CO oxidation. \ce{Au1}/\ce{CeO2}
has been reported\cite{Au1CeO2} and the geometry optimizations are
implemented with VASP, as shown in Fig.\ref{fig:Au1CeO2-CO}. CO can be
regarded as ACP detector in this example. It's worth mentioning that
this system is a periodic system. So the energy formula of periodic
system should be used. The VSEPR-SOPT result is shown in
Fig.\ref{fig:CeO2-Au-vsepr-sopt}. And the ACP acquired agrees well with
the position of carbon mentioned in the article. Angle difference is
\(13.19^\circ\), which is a really good result, Showing the validity of
the method in heterogeneous systems.

\section{SUMMARY AND DISCUSSION}

In this article, we purposed an efficient method for locating atom
connecting positions(ACPs). The basic idea is to substitute chemical
insight from VSEPR theory to a constrained geometry optimization
problem. With the help of spherical optimization(SOPT), the chosen
energy function and sampling method, all ACPs are acquirable and satisfy
chemical insight basically. Several typical systems are testified,
including small molecules, simple clusters, complicated molecular system
and heterogeneous system. In all systems except \ce{CH2O}@O, chemical
insight is satisfied rigorously, while for \ce{CH2O}@O, chemical insight
is partially satisfied as well, but the result is still acceptable. The
angle differences are less than \(20^\circ\), which is astonishingly
small. These examples demonstrate the validity and robustness of the
method. And the method can be useful in reaction searching and reaction
network construction.

\begin{acknowledgement}
This work was financially supported by the National Natural Science
Foundation of China (Grant Nos. 21590792, 91426302, and 2143005) to J.L.
and the Thousand Talents Plan for Young Scholars to H.X. The
calculations were performed using the supercomputers at the
Computational Chemistry Laboratory of Department of Chemistry under
Tsinghua Xuetang Talents Program. We thank Biao Yang from Tsinghua
University for providing the structure of \ce{Au6} cluster.
\end{acknowledgement}


\bibliography{article}

%

\end{document}